\documentclass{article}
\usepackage{spconf,amsmath,graphicx}
\usepackage{algorithm}
\usepackage{algorithmicx}
\usepackage{algpseudocode}
\usepackage{booktabs}
\usepackage{multirow}
\usepackage{color}
\usepackage{makecell}
\usepackage{subcaption}
\usepackage{url}

\usepackage{verbatim}
\usepackage{graphicx}
\usepackage{color}
\usepackage{url}
\usepackage{hyperref}

\title{LP-BFGS attack: An adversarial attack based on the Hessian with limited pixels}
\name{Jiebao Zhang$^1$, Wenhua Qian$^{1*}$,  Rencan Nie$^{1}$, Jinde Cao$^2$, Dan Xu$^1$
	\thanks{* Corresponding author: Wenhua Qian. Email: whqian@ynu.edu.cn. }
	\thanks{This work was supported by the Research Foundation of Yunnan Province No.202002AD08001, 202001BB050043,
		2019FA044, National Natural Science Foundation of China under Grants No.62162065, Provincial Foundation for Leaders of Disciplines in Science and Technology No.2019HB121, in part by the Postgraduate Research and Innovation Foundation of Yunnan University (No.2021Y281, No.2021Z078), and in part by the Postgraduate Practice and Innovation Foundation of Yunnan University (No.2021Y179, No.2021Y171).}
}
\address{$^1$School of Information Science and Engineering, Yunnan University, Kunming 650500, China  \\ 
	$^2$School of Mathematics, Southeast University, Nanjing 210096, China
}

\begin{document}
	%
	\maketitle
	\begin{abstract}
Deep neural networks are vulnerable to adversarial attacks. Most $L_{0}$-norm based white-box attacks craft perturbations by the gradient of models to the input. Since the computation cost and memory limitation of calculating the Hessian matrix, the application of Hessian or approximate Hessian in white-box attacks is gradually shelved.
In this work, we note that the sparsity requirement on perturbations naturally lends itself to the usage of Hessian information. We study the attack performance and computation cost of the attack method based on the Hessian with a limited number of perturbation pixels. 
Specifically, we propose the Limited Pixel BFGS (LP-BFGS) attack method by incorporating the perturbation pixel selection strategy and the BFGS algorithm. 
Pixels with top-k attribution scores calculated by the Integrated Gradient method are regarded as optimization variables of the LP-BFGS attack.
Experimental results across different networks and datasets demonstrate that our approach has comparable attack ability with reasonable computation in different numbers of perturbation pixels compared with existing solutions.
	\end{abstract}
		\section{Introduction}
	\label{sect:intro}
	Deep Neural Networks (DNNs) have surpassing performance on the image classification task \cite{image_classification_survey}.
	However, researchers have found that DNNs are highly susceptible to small malicious perturbations crafted by adversaries \cite{evasion_attack,L-BFGS_attack}.
	Specifically, malicious perturbations in original examples can significantly harm the performance of DNNs.
	DNNs are therefore untrustworthy for security-sensitive tasks.
	Many adversarial attack methods have been proposed to seek perturbations according to the unique properties of DNNs and optimization techniques.
	\par
	Depending on the attacker's knowledge of the target model, adversarial attacks can be divided into two categories: white-box attacks and black-box attacks.
	White-box attacks assume that attackers have detailed information about the target model (\textit{e.g.}, the training data, model structure, and model weight), and they can be further classified into optimization-based attacks \cite{evasion_attack,L-BFGS_attack,C&W}, single-step attacks \cite{FGSM,Random-FGSM}, and iterative attacks \cite{PGD,I-FGSM,MI-FGSM,DeepFool,UAP,JSMA,SparseFool}.
	Optimization-based attacks formulate finding the optimal perturbation as a box-constrained optimization problem. 
	Szegedy \textit{et al.} use the quasi-Newton method, Limited-memory BFGS method \cite{L-BFGS,L-BFGS-2}, to solve the box-constrained problem, called L-BFGS attack \cite{L-BFGS_attack}.
	Compared with the L-BFGS attack, the C\&W \cite{C&W} attack uses variable substitution to bypass the box constraint and uses a more efficient objective function.
	Furthermore, it uses the Adam optimizer \cite{Adam} to find the optimal perturbation.
	Single-step attacks are simple and efficient and can alleviate the high computation cost caused by optimization-based attacks. 
	Since the model is assumed to be locally linear, perturbations in single-step attacks are added directly along the gradient \cite{FGSM,Random-FGSM}.
	Iterative attacks add perturbations in multiple steps, achieving a tradeoff between the computation and the attack performance.
	Black-box attacks mean that attackers have little information about the architecture and parameters of the target model.
	Compared with white-box attacks, they also can achieve an equivalent attack by querying the output (\textit{e.g.}, the confidence score or final decision) of the model \cite{OPA,ZOO_attack,Boundary_attack,Sparse-RS}.
	\par
	\begin{figure*}
		\centering
		\includegraphics[width= 1.0 \textwidth]{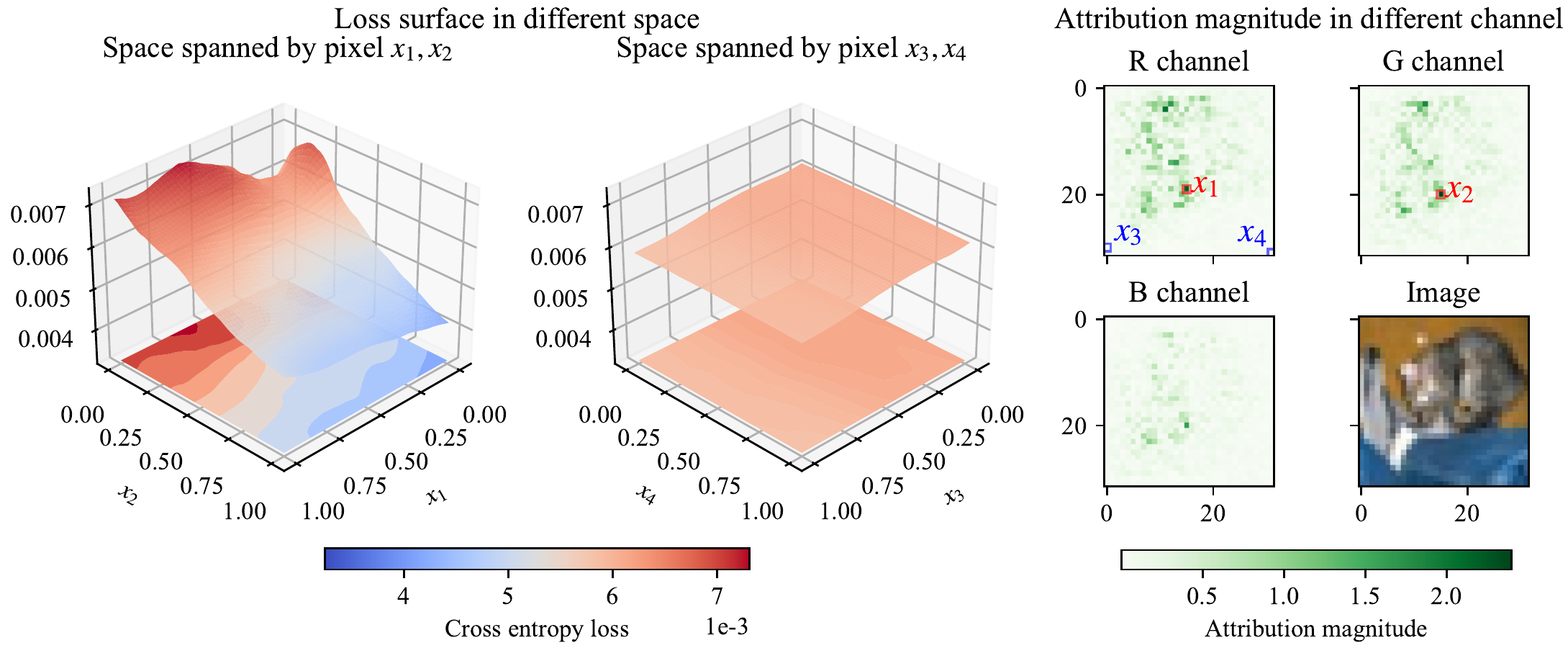}
		\caption{The illustration that pixels in different locations play a different role in the loss.}
		\label{fig:pixel-effect-opt}
	\end{figure*}
	\begin{figure*}
		\centering
		\includegraphics[width= 0.7 \textwidth]{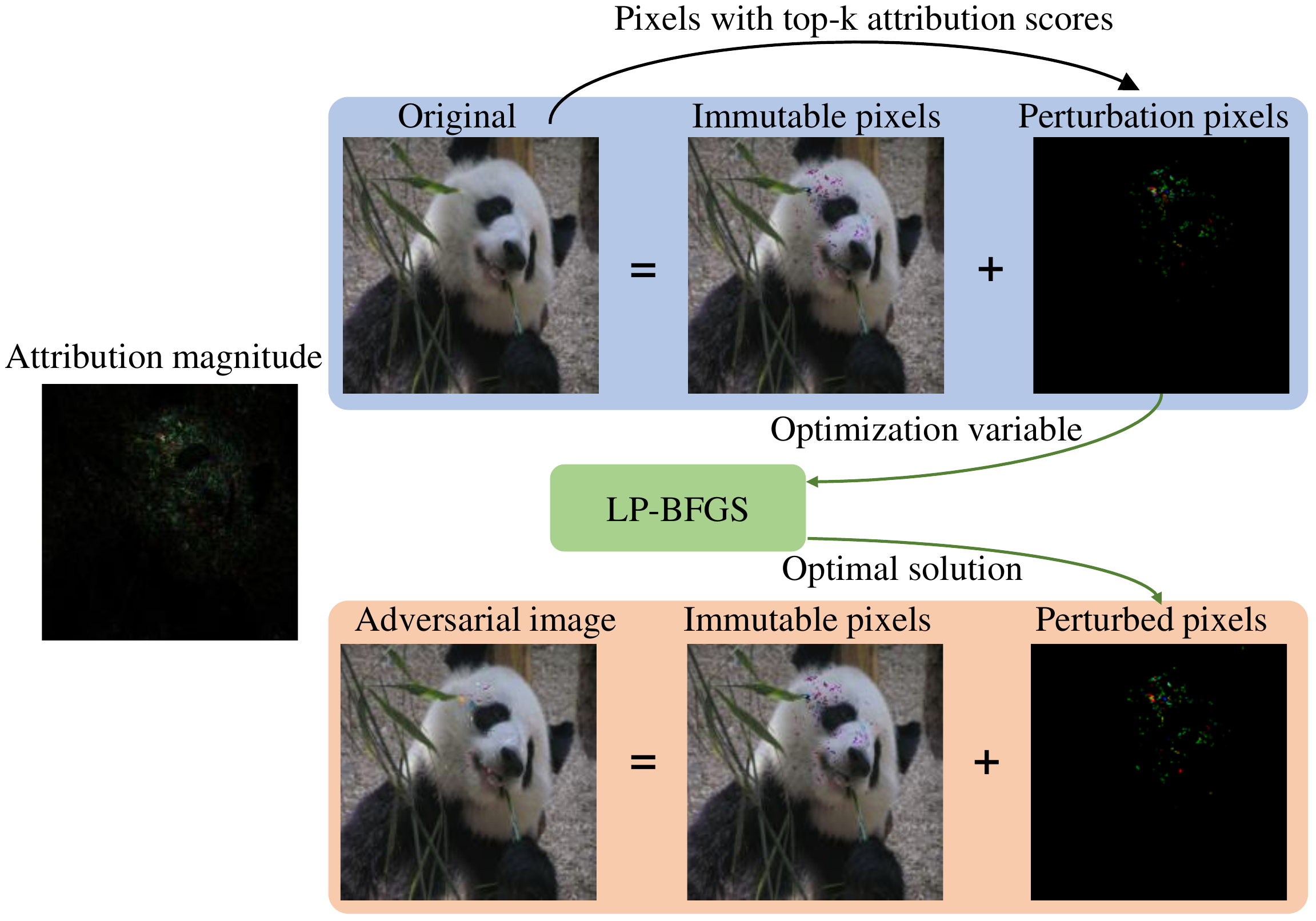}
		\caption{The attack framework of LP-BFGS. Pixels with top-k attribution scores calculated by the Integrated Gradient method are selected as optimization variables of LP-BFGS. The adversarial image combines immutable pixels with the optimal perturbation obtained by LP-BFGS.}
		\label{fig:perturbations-generated-by-lp-bfgs}
	\end{figure*}
	Most existing white-box and black-box attacks have in common the tendency to indiscriminately perturb all pixels of images. But some research has shown that attackers can achieve strong attack effects by perturbing certain regions or pixels of original images. 
	JSMA \cite{JSMA} selects one or more pixels that play an important role in the model’s prediction for modification at the current iteration. 
	C\&W \cite{C&W} do iterative executions of the $L_2$ distance attack to obtain perturbations with minimal $L_0$ distance. 
	SparseFool \cite{SparseFool} exploits the low mean curvature of the decision boundary to control the sparsity of the perturbations.
	OPA \cite{OPA} is a score-based attack that is based on the differential evolution algorithm to generate adversarial perturbations in the black-box scenario.
	Sparse-RS \cite{Sparse-RS} is also a score-based black-box attack based on the random search.
	Furthermore, some sparse perturbations can be deployed in the physical world \cite{attack_road_sign, attack_face_recognition,IS_10}.
	\par
	The sparsity of perturbations is indeed visible but does not alter the semantic content \cite{Sparse-RS}. Moreover, the demand for sparsity paves the way for the usage of second-order gradient information.
	A sizeable image will commonly bring in the high-dimension optimization variable in the attack process. The second-order gradient information of the loss function w.r.t. the original image, namely the Hessian matrix, requires expensive computational cost and memory budget. For example, for a size $3 \times 256 \times 256$ image, the size of the loss function's Hessian matrix on it is $(3 \cdot {2^{16}}) \times (3 \cdot {2^{16}})$. This matrix would require 144 gigabytes of memory space simultaneously, assuming each element in the matrix is a floating-point data represented by 4 bytes. This makes it difficult to use second-order gradient information for most attack algorithms, and the high dimensionality of the optimization variables is a key factor hindering the use of Hessian or approximate Hessian information for attacks. 
	\par
	However, when the number of perturbed pixels is limited, which pixels are more conducive to the attack requires a lot of careful consideration. Perturbing pixels at different locations may produce different attack effects. Fig.~\ref{fig:pixel-effect-opt} shows the modified pixels at different locations have different impacts on the model's loss value.
	\par
	In this work, we propose a limited-pixel BFGS attack (LP-BFGS) method incorporating attribution scores of pixels and approximation Hessian information. Specifically, for an image, we use the integrated gradient algorithm to compute the attribution score of each pixel with respect to the model's decision about the true label. We select some pixels with high attribution scores as perturbation pixels to reduce the dimensionality of the optimized variables and find adversarial examples under the guidance of approximate Hessian information, as shown in Fig.~\ref{fig:perturbations-generated-by-lp-bfgs}. The main contributions of this work are as follows:
	\begin{itemize}
		\item{We propose the LP-BFGS attack method by incorporating the perturbation pixel selection strategy and the BFGS algorithm. LP-BFGS only perturbs some pixels with the guidance of the Hessian.}
		\item{We investigate the effect of loss functions and perturbation pixel numbers on the performance of the LP-BFGS attack and study the time cost of the LP-BFGS attack family.}
		\item{We conduct experiments across various datasets and models to verify that the LP-BFGS attack can achieve a comparable attack with an acceptable computation cost.}
	\end{itemize}
	\begin{figure*}[!t]
		\centering
		\includegraphics[width=1.0\linewidth]{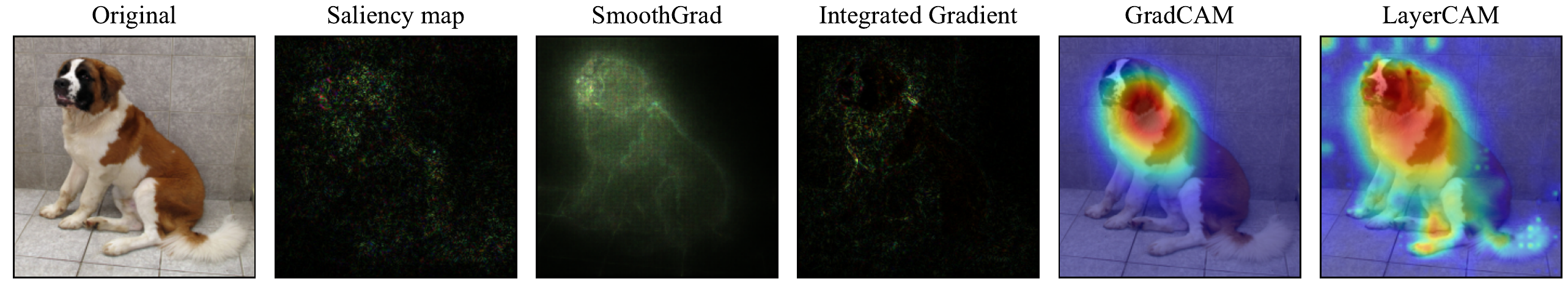}
		\includegraphics[width=1.0\linewidth]{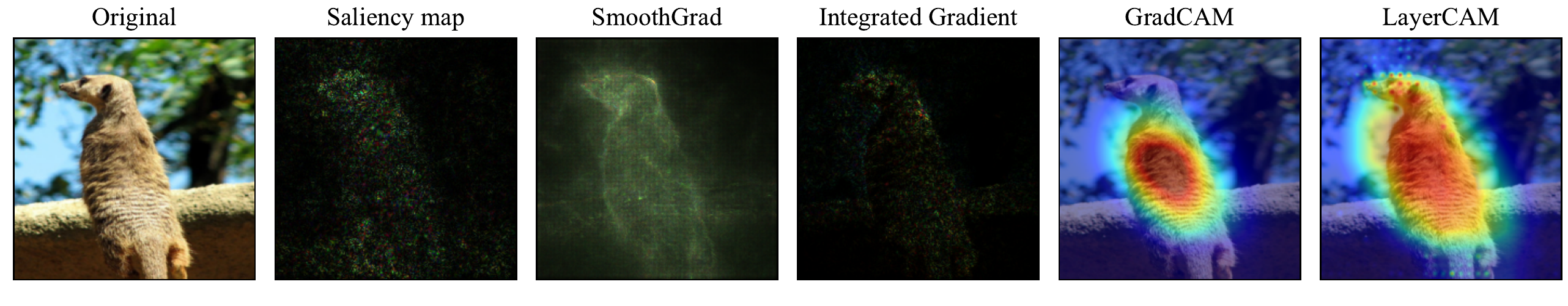}
		\caption{The attribution results of Saliency map \cite{Saliency_map}, SmoothGrad \cite{SmoothGrad}, Integrated Gradient \cite{integrated_gradient}, GradCAM \cite{Grad_CAM} and LayerCAM \cite{LayerCAM} methods.}
		\label{fig:different-attributions}
	\end{figure*}
	
	\section{Related Work}
	
	\subsection{Adversarial attacks}
	Various adversarial attacks have been proposed to seek perturbations according to the exclusive attributes of DNNs and optimization techniques. Some of them investigate the robustness of models against sparse perturbations. 
	OPA selects candidate solutions generated by differential evolution as the adversarial examples \cite{OPA}. Sparse-RS applies components to be perturbed and the corresponding perturbation values to form the adversarial input \cite{Sparse-RS}. Some white-box attacks have been proposed.
	\par
	\textbf{L-BFGS attack.} L-BFGS formulates finding an optimal perturbation as a box constraint optimization problem:
	\begin{equation}
		\underset{r}{\min} \; c\|r\|_p+ f(x+r, t) \quad \text{s.t.} \; x+r\in[0,1]^n,
	\end{equation}
	where $r$ denotes the perturbation vector, $x$ is original input, $t$ is the target label, and $n$ is size of the image $x$. $f$ is the loss function customized by attackers, \textit{e.g.}, the cross entropy. The constant $c$ finds a compromise between the perturbation magnitude and the attack performance.
	The L-BFGS attack uses the second-order quasi-Newton method \cite{L-BFGS,L-BFGS-2} to solve this problem.
	\par
	\textbf{FGSM.} FGSM directly adds perturbations along the gradient direction of the model to the original examples since models are assumed to be locally linear. Therefore, it would avoid the high computation costs caused by the optimization procedure in a straight and effective manner:
	\begin{equation}
		\hat{x}= x + \epsilon \cdot \text{sign}[\nabla_x J(\theta,x,y)]
	\end{equation}
	where $\hat{x}$ denotes the adversarial example, $\epsilon$ is the magnitude of the perturbation, $J$ is the loss function, and $\theta$ is the parameter of the target model. 
	\par
	\textbf{JSMA.} JSMA is a $L_0$-norm based white box targeted attack method \cite{JSMA}. This attack is based on the greedy strategy that one or more pixels with a high impact on the prediction are selected for modification in each iteration. Concretely, The JSMA attack calculates the attack saliency map by the Jacobian matrix of the model, which reflects the importance of pixels in the model prediction. The calculation method for the saliency map $S$ is as follows:
	\begin{equation}
		S{(x,t)_i} = \left\{ 
		\begin{aligned}
			{0,\;{\text{if}}\frac{{\partial Z{{(x)}_t}}}{{\partial {x_i}}} < 0\;{\text{or}}\;\sum\limits_{j \ne t} {\frac{{\partial Z{{(x)}_t}}}{{\partial {x_i}}}}  > 0} \\ 
			{\left( {\frac{{\partial Z{{(x)}_t}}}{{\partial {x_i}}}} \right)\left| {\sum\limits_{j \ne t} {\frac{{\partial Z{{(x)}_j}}}{{\partial {x_i}}}}} \right|{\text{,}}\;{\text{otherwise}}} 
		\end{aligned}
		\right.
	\end{equation}
	where $x$ denotes the input, $t$ is the target label, and $Z(x)$ is the logits.
	\par
	\textbf{C\&W.} Compared with the L-BFGS attack, the C\&W attack uses variable substitution to bypass the box constraint, replace the objective with a more powerful one, and uses the Adam optimizer \cite{Adam} to solve the optimization problem \cite{C&W}, the expression as follows:
	\begin{equation}
		\underset{r}{\min} \; \| r \|_{p}+c \cdot f(x+ r) \quad \text {s.t.} \; x+ r \in[0,1]^{n},
		\label{eq: cw_attack}
	\end{equation}
	where $f$ is the C\&W loss function and the detailed version is as follows:
	\begin{equation}
		f ( \cdot )= \max \left( \max \left\{ Z( \cdot )_{i, i \neq t} \right\} - Z( \cdot )_{t},-\kappa \right),
	\end{equation}
	where $Z(\cdot)$ represents the logits output of the target model, $t$ is the target label that attackers expect the model to predict, and $\kappa \geq 0$ is a hyperparameter for attack transferability. A new variable $w$ is introduced to substitute $r$ and the relationship between them is $r =\frac{1}{2}(\tanh (w)+1)-x$, such that $x+r = \frac{1}{2} (\tanh(w)+1)$, which always resides in the range of $[0,1]$ in the optimization process.
	\par
	\textbf{SparseFool.} SparseFool exploits the low mean curvature of the decision boundary to generate sparse perturbations \cite{SparseFool}.
	In each iteration, SpareFool will obtain the boundary point $x_B$ located on the decision boundary with the help of the $L_2$ attack DeepFool \cite{DeepFool} and estimate the normal vector $w$ to the boundary at the datapoint $x_B$. 
	The sparse perturbation is found by a subprocedure, called LinearSolver, which iteratively projects perturbations only towards one single coordinate of the normal vector $w$  until the adversarial point reaches the boundary or there is no space to search.
	
	\begin{figure*}[!t]
		\centering
		\includegraphics[width= 0.6 \textwidth]{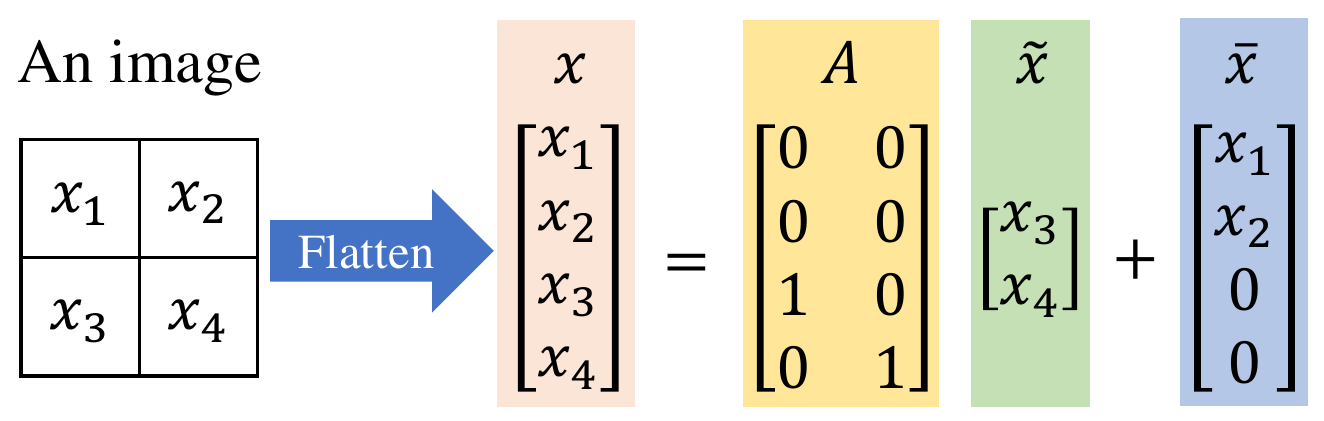}
		\caption{The illustration that a $2 \times 2$ image endures procedures described in Algorithm~\ref{algo:pixel-selector}. $x_3$ and $x_4$ are selected as the perturbed pixel.}
		\label{fig:pixel_selector}
	\end{figure*}
	\subsection{Attribution Score}

	Some pioneers’ works identify the role of pixels or regions in images in the final decision of DNNs \cite{Saliency_map,integrated_gradient,DeepLIFT,LIME,CAM,Grad_CAM,LayerCAM}. 
	\par
	Saliency Map approximates the class score function with a first-order Taylor expansion in the neighborhood of an image \cite{Saliency_map}. The magnitude of the derivative of the class score function with respect to the image indicates which pixels need to be changed the least to affect the class score the most \cite{Saliency_map}.
	To reduce visual noise, SmoothGrad samples similar images by adding noise to the image, then take the average of the resulting sensitivity maps for each sampled image \cite{SmoothGrad}.
	The Integrated Gradient (IG) method is designed under the guidance of the axioms-\textit{Sensitivity} and \textit{Implementation Invariance} that attribution methods ought to satisfy \cite{integrated_gradient}.
	The implementation detail is as follows:
	\begin{equation}
		\text{IG}_i(x)=\left(x_i-x_i^{\prime}\right) \times \int_{\alpha=0}^1 
		\frac{\partial F\left(x^{\prime}+\alpha\times\left(x-x^{\prime}\right)\right)}
		{\partial x_i}d \alpha,
		\label{eq:IG}
	\end{equation}
	where the function $F$ represents the deep neural network, $x$ represents the input, and $x^{\prime}$ represents the baseline input. The attribution of the $i$-th component can be regarded as the accumulation of all gradients on the path from the baseline $x^{\prime}$ to the input $x$.
	\par
	The Class Activation Maps (CAMs) are obtained by multiplying each feature map by its weight and then performing a summation on all weighted feature maps \cite{LayerCAM}.
	CAM \cite{CAM} obtained the weights from the retrained fully-connected layer.
	Grad-CAM \cite{Grad_CAM} averages the gradients of the target class with respect to each feature map as its weight. LayerCAM generates reliable class activation maps from different layers of CNN \cite{LayerCAM}.
	\par
	Class activation maps visualize the regions in which models are interested in a coarse-grained manner. In this work, as shown in Fig.~\ref{fig:different-attributions}, we focus on the role of the fine-grained feature, \textit{e.g.}, pixels in different channels. We take in attribution scores generated by IG as expert knowledge.
	
	\section{Method}

	\subsection{Problem Formulation}
	\label{sect:problem_formulation}
	An image with the channel $C$, height $H$, and width $W$, is denoted as a vector $x \in R ^ {C \times H \times W}$, and the corresponding label is $y$.
	A deep neural image $M$-classifier can be regarded as a function $g$: $R^{C \times H \times W} \rightarrow R^M$.
	In this work, we use $Z$ to denote logits \cite{C&W} of a model and $\bar{Z}$ to represent the probability.
	The prediction of $g$ for the image $x$ is $\arg \max_i g(x)_i$.
	Given the distance function $d(\cdot,\cdot)$ and $\varepsilon>0$, an adversarial example $\hat{x}$ satisfies the following formula:
	\begin{equation}
		\arg \max_i \; g(\hat{x})_i \neq y, \quad d(x,\hat{x}) \leq \varepsilon.
		\label{eq:adversarial_example}
	\end{equation}
	The adversarial attack is to deal with the intractable problem \eqref{eq:adversarial_example} and seek the adversarial example.
	In general, the objective function in \eqref{eq:adversarial_example} is relaxed to minimize the loss $L$ customized by attackers, and the distance function is the Euclidean distance. The adversarial attack can be formulated as follows:
	\begin{equation}
		\min \; L(g(\hat{x}),y) \quad
		\text{s.t.} \|\hat{x}-x\| \leq \varepsilon.
		\label{eq:adversarial_attack}
	\end{equation}
	
	\subsection{Pixel Selector}
	\label{sect:pixel_selector}

	A majority of adversarial attacks usually perturb all pixels of an image.
	In this work, we limit the number of pixels that can be perturbed according to the magnitude of the attribution score of pixels with respect to the final decision.
	Concretely, for an image $x$, we take into consideration pixels in different channels and use IG \cite{integrated_gradient} to assign the attribution magnitude to each pixel with respect to the label $y$.
	Pixels whose attribution scores are top-$K$ are selected as perturbed pixels and denoted as a vector $\tilde{x} \in R^K$ and $K << C \times H \times W$.
	When the pixels that can be perturbed in the original image $x$ are removed, the remaining immutable pixels are recorded as $\bar{x} \in R^{C \times H \times W}$.
	A matrix $A$ records the position map between $\tilde{x}$ and $x$.
	The relationship among $A$, $\tilde{x}$, and $\bar{x}$ is as follows:
	\begin{equation}
		x = A \tilde{x} + \bar{x}.
		\label{eq:image_decomposition}
	\end{equation}
	We can use it to reconstruct the adversarial image after obtaining the optimal perturbation, as shown in Fig.~\ref{fig:pixel_selector}.
	The size of matrix $A$ is equal to $(C \times H \times W) \times K$, which is proportional to the number of perturbed pixels.
	\begin{algorithm}[!t]
		\caption{Pixel Selector}
		\label{algo:pixel-selector}
		\begin{algorithmic}[1]
			\Require{Model $F(x)$, image $x$, corresponding label $y$, dimension $N$, perturbed pixel number $K$.}
			
			\Ensure{Position map $A$, perturbed pixel $\tilde{x}$, immutable pixel $\bar{x}$.}
			\State{Initialize $A \leftarrow 0, \tilde{x} \leftarrow 0, \bar{x} \leftarrow 0$}
			\State{$S \leftarrow$ Record the attribution magnitude by \eqref{eq:IG}}
			\State{$D \leftarrow$ Obtain the index of the top-$K$ elements in $S$}
			\For{$i$ \textbf{in} $1,2,\cdots,K$}
			\State{$\tilde{x}_i \leftarrow x_{D_i}$}
			\State{$A_{D_i,i} \leftarrow 1$}
			\EndFor 
			\State{$\bar{x} \leftarrow x-A \tilde{x}$} \\
			\Return {$A,\tilde{x},\bar{x}$}
		\end{algorithmic}
	\end{algorithm}
	\begin{algorithm}[!t]
		\caption{Limited Pixel BFGS Attack}
		\label{algo:LP-BFGS}
		\begin{algorithmic}[1]
			\Require{Image $x$, corresponding label $y$, loss function $L\left(\tilde{w},x,y\right)$.}
			\Require{Convergence tolerance $\epsilon > 0$, iteration $T>0$.}
			\Ensure{Adversarial image $\hat{x}$.}
			\State {Obtain $A$, $\tilde{x}$, and $\bar{x}$ by Algorithm~\ref{algo:pixel-selector}}
			\State {Initialize $k \leftarrow 0, \; H^k \leftarrow I$}
			\State {$\tilde{w}^k \leftarrow \text{atanh}(2 \tilde{x} -1)$}
			\While{$\|\nabla_{\tilde{w}} L^k \| > \epsilon$ \textbf{or} $k<T$}
			\State {$d^k \leftarrow -H^k \nabla_{\tilde{w}} L^k$}
			\State {$\tilde{w}^{k+1} \leftarrow \tilde{w}^k + \alpha^k d^{k}$ where $\alpha^k$ satisfies the Wolfe condition.}
			\State{$s^k \leftarrow \tilde{w}^{k+1}-\tilde{w}^k$}
			\State{$y^k \leftarrow \nabla_{\tilde{w}} L^{k+1} - \nabla_{\tilde{w}} L^k$}
			\State{Compute $H^{k+1}$ by \eqref{eq:inverse_Hessian_update_strategy}}
			\State{$k \leftarrow k+1$}
			\EndWhile
			\State{$\hat{w} \leftarrow \tilde{w}^k$}
			\State{Obtain the adversarial image $\hat{x}$ by \eqref{eq:image_from_wp}} \\
			\Return {$\hat{x}$}
		\end{algorithmic}
	\end{algorithm}
	\subsection{General BFGS Method.}
	For the objective function $f(x)$, Newton methods use the Taylor series to make a second-order approximation of the function at the current iterate $x^k$ and find minimal value:
	\begin{equation}
		m^k(p) = f^k + \nabla {(f^k)}^T p + \frac{1}{2}p^T B^k p,
		\label{eq:taylor_series}
	\end{equation}
	where $m^k(p)$ denotes the approximation function, $f^k = f(x^k)$, and $B^k$ is an $n \times n$ symmetric positive definite matrix.
	The search direction $d^{k}$ is:
	\begin{equation}
		d^{k} = -{(B^k)}^{-1} \nabla f^k,
		\label{eq:search_direction}
	\end{equation}
	and the new iteration is:
	\begin{equation}
		x^{k+1} = x^k +\alpha^k d^{k},
	\end{equation}
	where $\alpha^k$ denotes the step length.
	Notably, the computation and memory costs of the Hessian matrix in the optimization are considerable, when the variable has a large size.
	\par
	Instead of computing the Hessian, the popular quasi-Newton method, the BFGS method, revises the approximate Hessian $B^k$ (or inverse Hessian $H^k$) in $k$-th iteration by the most recently observed information about the objective function \cite{quasi_Newton_in_numerical_optimization}.
	To make the approximate Hessian as close as possible to the real Hessian, the approximate Hessian $B^k$ should satisfy the \textit{secant equation}:
	\begin{equation}
		B^{k+1} s^k = y^k,
		\label{eq:secant_equation}
	\end{equation}
	where $s^k = x^{k+1} - x^k$ and $y^k = \nabla f^{k+1} -\nabla f^k$. 
	By premultiplying \eqref{eq:secant_equation} by $s^k$, we have $(s^k)^T B^{k+1} s^k = (s^k)^T y^k$. 
	$B^{k+1}$ is positive definite, we have:
	\begin{equation}
		{(s^k)}^T y^k >0,
		\label{eq:curvature_condition}
	\end{equation}
	which is \textit{curvature condition}. In nonconvex functions, the curvature condition is guaranteed by imposing the Wolfe condition on the line search. 
	The common update strategy of $B^{k+1}$ is as follows:
	\begin{equation}
		{B^{k + 1}} = {B^k} + \frac{{{y^k}{{\left( {{y^k}} \right)}^{T}}}}{{{{\left( {{s^k}} \right)}^{T}}{y^k}}} - \frac{{{B^k}{s^k}{{\left( {{B^k}{s^k}} \right)}^{T}}}}{{{{\left( {{s^k}} \right)}^{T}}{B^k}{s^k}}}.
	\end{equation}
	Moreover, to alleviate the computation of solving equations in \eqref{eq:search_direction}, the inverse Hessian approximation $H^k$ is updated as follows: 
	\begin{equation}
		H^{k+1}=\left(I-\rho^k s^k {(y^k)}^T\right) H^k\left(I-\rho^k y^k {(s^k)}^T\right)+\rho^k s^k {(s^k)}^T.
		\label{eq:inverse_Hessian_update_strategy}
	\end{equation}
	where $\rho_{k} = \frac{1}{{(y^k)}^T s^k}$.
	
	\begin{figure*}[!t]
		\centering
		\subfloat[Res-20 on CIFAR-10 with $k=25,\lambda=1.0$]{
			\centering
			\begin{minipage}[c]{0.9 \textwidth}
				\includegraphics[width=1 \textwidth]{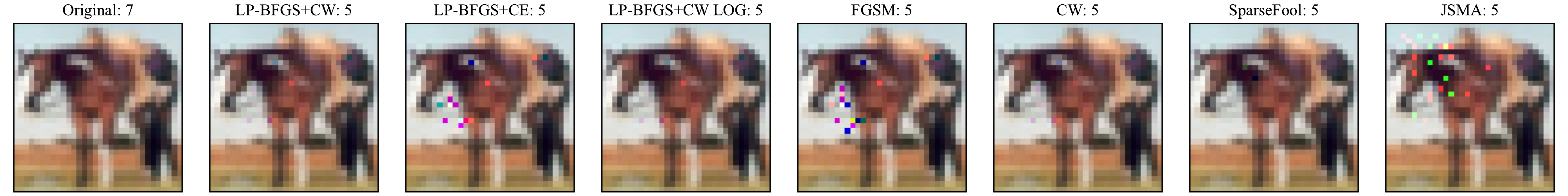}
				\includegraphics[width=1 \textwidth]{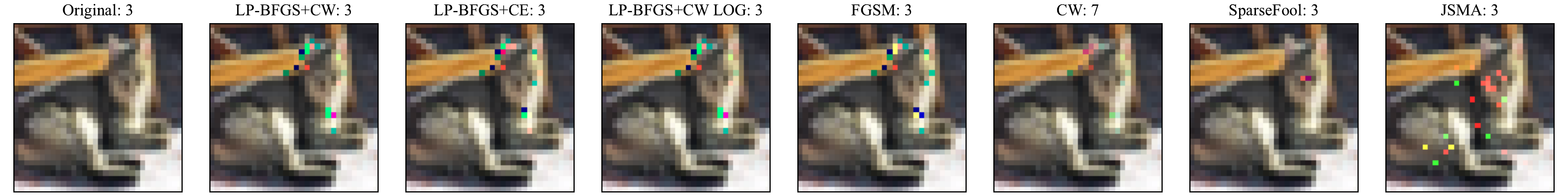}
			\end{minipage}
		}
		\\
		\subfloat[Res-20 on CIFAR-10 with $k=40,\lambda=1.6$]{
			\centering
			\begin{minipage}[c]{0.9 \textwidth}
				\includegraphics[width=1 \textwidth]{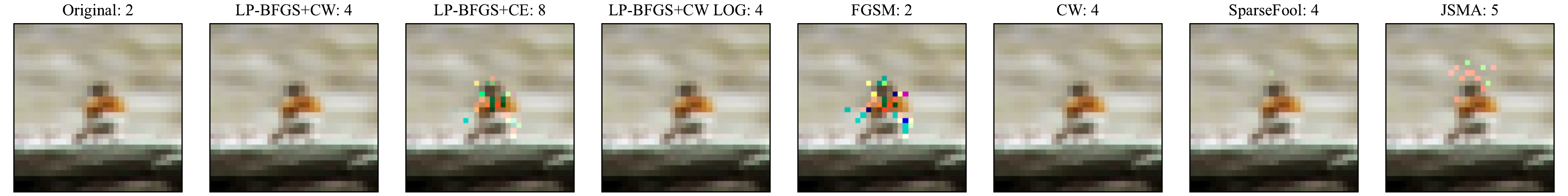}
				\includegraphics[width=1 \textwidth]{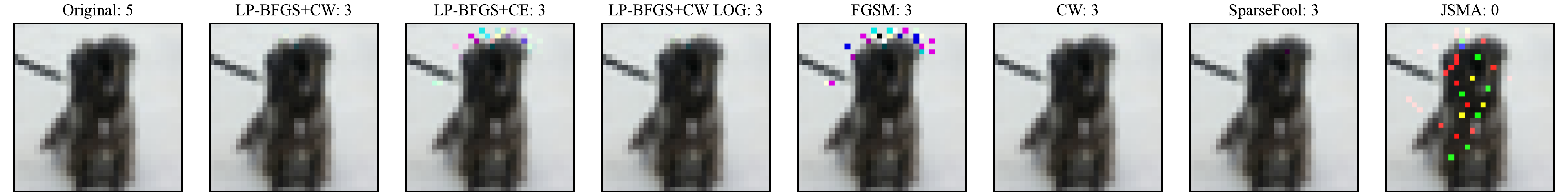}
			\end{minipage}
		}
		\\
		\subfloat[Res-34 on ImageNet with $k=400$]{
			\centering
			\begin{minipage}[c]{0.9 \textwidth}
				\includegraphics[width=1 \textwidth]{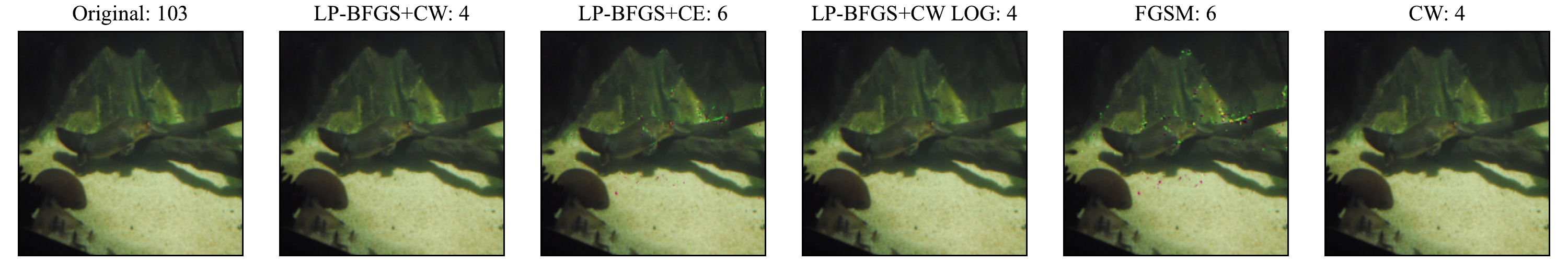}
				\includegraphics[width=1 \textwidth]{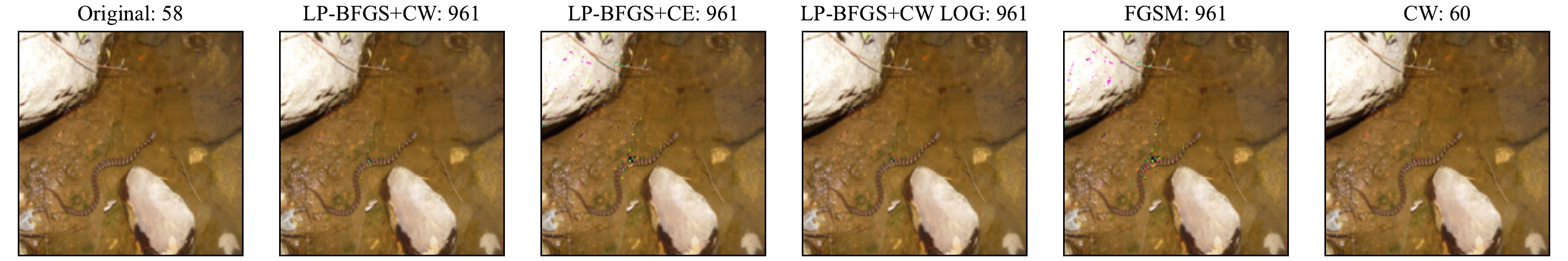}
			\end{minipage}
		}
		\\
		\subfloat[Res-34 on ImageNet with $k=1000$]{
			\centering
			\begin{minipage}[c]{0.9 \textwidth}
				\includegraphics[width=1 \textwidth]{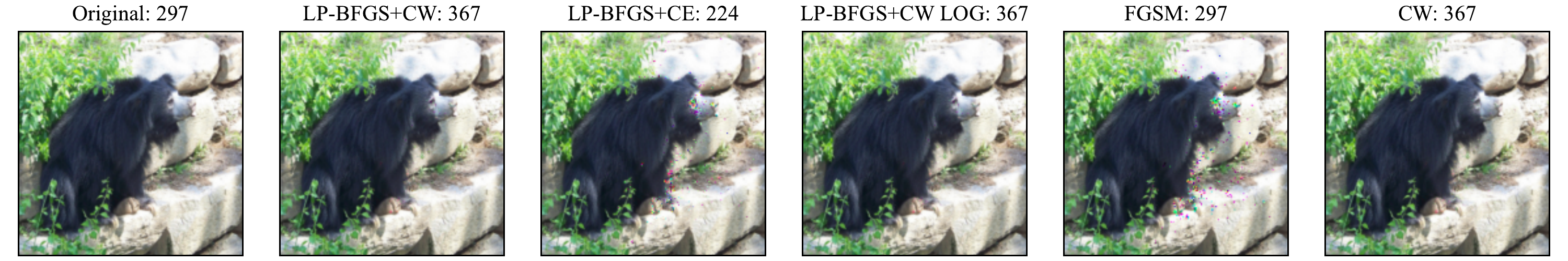}
				\includegraphics[width=1 \textwidth]{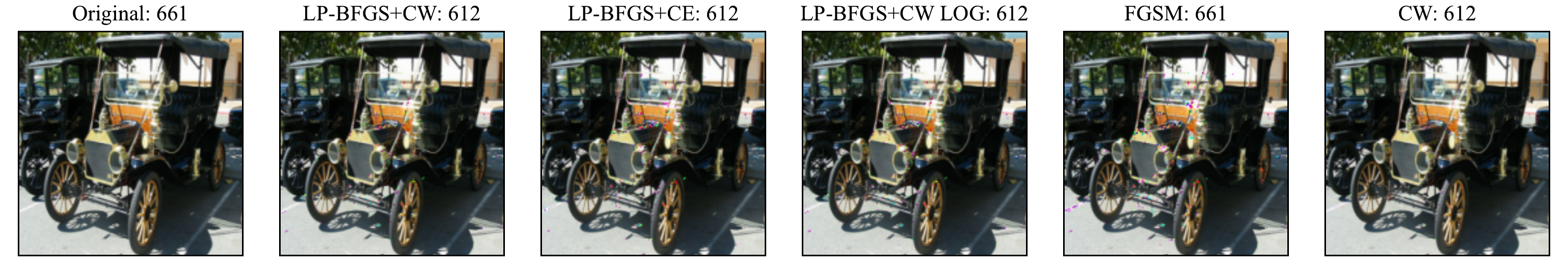}
			\end{minipage}
		}
		\caption{Adversarial examples on CIFAR-10 and ImageNet with different numbers of perturbed pixels. The title on each image indicates the attack method and label given by the target model.}
		\label{fig:adv-images}
	\end{figure*}
	
	\subsection{Limited Pixel BFGS Attack}
	\label{sect:lp-bfgs}
	
	In Section \ref{sect:problem_formulation}, the adversarial attack is formulated as an optimization problem with a constraint.
	In \cite{L-BFGS_attack}, the adversarial attack is formulated as a box-constrained problem. This problem is then reformulated as an unconstrained problem by the variable substitution \cite{C&W}.
	In this paper, we absorb the pioneers' scheme and incorporate the pixel selector mentioned in Section \ref{sect:pixel_selector}.
	We propose a novel $L_0$-norm based white-box attack method, LP-BFGS. Compared with C\&W, we utilize second-order information. Furthermore, we do not use Limited Memory BFGS, as we constrain the number of perturbed pixels and greatly reduce the size of the Hessian matrix. BFGS calculates and stores the full Hessian matrix, which provides a more accurate estimate of the optimization landscape than L-BFGS. Different from JSMA, LP-BFGS does not calculate the saliency map during each iteration but rather determines the space of perturbations at the beginning of the attack. The procedure of LP-BFGS is expressed in Algorithm \ref{algo:LP-BFGS}.
	\par
	LP-BFGS introduces a new variable $\tilde{w} \in (-\infty,+\infty)^K$ to bypass the constraint $\tilde{x} \in [0,1]^K$:
	\begin{equation}
		\tilde{w} =\text{atanh}(2 \tilde{x} -1).
	\end{equation}
	In this way, we convert the optimization problem with the variable $x$ into the one with the variable $\tilde{w}$, dramatically reducing the optimization variable's dimension and relieving the computation burden and memory consumption.
	This proposed strategy also bypasses the box constraint.
	So we can solve the unconstrained optimization problem to find optimal adversarial examples by the BFGS optimizer:
	\begin{equation}
		\hat{w} = \underset{\tilde{w}}{\min} \;  L\left(\tilde{w}, x, y\right),
		\label{eq:}
	\end{equation}
	where $\hat{w}$ is the optimization result.
	The corresponding adversarial image would be reconstructed by: 
	\begin{equation}
		\hat{x} = R(\hat{w})= A \left(\frac{1}{2} (\tanh(\hat{w}) + 1)\right) + \bar{x}.
		\label{eq:image_from_wp}
	\end{equation}
	Moreover, we present three objective functions of the LP-BFGS attack in the untargeted attack scenario, which are mostly the same except for the $\hat{L}$:
	\begin{equation}
		L = \|R(\tilde{w})-x\| + c \cdot \hat{L}.
	\end{equation}
	There are three kinds of formulas of $\hat{L}$:
	\begin{equation}
		\hat{L}_{\text{CE}} = -(-\log \bar{Z} ( \tilde{w} )_{y}) \label{eq:CE_loss}, 
	\end{equation}
	\vspace{-1em}
	\begin{equation}
		\hat{L}_{\text{CW}}  = \max  \left \{ \max  \left \{Z( \tilde{w} )_{y}-Z( \tilde{w} )_{i, i \neq y} \right \},-\kappa  \right \}, \label{eq:CW_loss}
	\end{equation}
	\vspace{-1em}
	\begin{equation}
		\hat{L}_{\text{CW LOG}}= \max  \left \{ \max  \left \{ \log \bar{Z} ( \tilde{w} )_{y}- \log \bar{Z}( \tilde{w} )_{i, i \neq y} \right \},-\kappa  \right \}, \label{eq:CW_log_loss} 
	\end{equation}
	where $Z( \tilde{w} )_{i} = g(R(\tilde{w}))_{i}$, $\bar{Z} ( \tilde{w} )_i = \text{softmax} \left(g(R(\tilde{w}))\right)_{i}$, the constant $c$ implicitly controls the balance between the perturbation magnitude and the classification loss, and $\kappa \geq 0$ is a hyperparameter for the attack transferability.
	
	\begin{table*}[!t]
		\centering
		\caption{The performance of LP-BFGS (with 3 loss functions), FGSM \cite{FGSM}, C\&W \cite{C&W}, SpareFool \cite{SparseFool} and  JSMA \cite{JSMA} on the CIFAR-10 dataset. The target models are Res-20 \cite{ResNet} and NIN \cite{NIN}.}
		\resizebox{\textwidth}{!}{
			\begin{tabular}{llcccccccccccc}
				\toprule
				& \multicolumn{7}{c}{Res-20}                            & \multicolumn{6}{c}{NIN} 
				\\
				\cmidrule(l){3-8} \cmidrule(l){9-14}  
				\multicolumn{1}{c}{Pixel / $\lambda$} & Attack & ASR (\%) & Confidence & \multicolumn{1}{c}{$L_0$} & \multicolumn{1}{c}{$L_1$} & \multicolumn{1}{c}{$L_2$} & \multicolumn{1}{c}{$L_{\infty}$} & ASR (\%) & Confidence & \multicolumn{1}{c}{$L_0$} & \multicolumn{1}{c}{$L_1$} & \multicolumn{1}{c}{$L_2$} & \multicolumn{1}{c}{$L_{\infty}$} 
				\\
				\cmidrule{1-14}
				& LP-BFGS+CW & 59.58 & 0.44  & 20    & 2.87  & 0.8   & 0.37  & 51.36  & 0.36  & 20.00  & 3.56  & 0.98  & 0.43  \\
				& LP-BFGS+CE & 56.01 & \textbf{0.81} & 20    & 6.76  & 1.8   & 0.74  & \textbf{52.10} & \textbf{0.54} & 20.00  & 8.12  & 2.09  & 0.80  \\
				& LP-BFGS+CW LOG & \textbf{59.69} & 0.44  & 20    & 2.86  & 0.8   & 0.37  & 50.99  & 0.36  & 20.00  & 3.51  & 0.97  & 0.43  \\
				\multicolumn{1}{c}{k=20} & FGSM  & 25.17 & 0.67  & 19.96 & 10.25 & 2.53  & 0.85  & 31.81  & 0.47  & 19.95  & 10.54  & 2.59  & 0.86  \\
				\multicolumn{1}{c}{$\lambda=1.0$} & C\&W  & 55.46 & 0.44  & 20    & 1.87  & 0.5   & 0.21  & 50.50  & 0.36  & 20.00  & 2.85  & 0.72  & 0.27  \\
				& SparseFool & 9.67  & 0.51  & 459.74 & 230.24 & 4.88  & 0.36  & 23.27  & 0.44  & 377.87  & 189.23  & 4.30  & 0.56  \\
				& JSMA  & 30.11 & 0.62  & 15.25 & 8.36  & 2.2   & 0.79  & 29.46  & 0.43  & 14.65  & 8.23  & 2.20  & 0.81  \\
				\midrule
				& LP-BFGS+CW & 84.26 & 0.44  & 40    & 4.58  & 0.92  & 0.33  & 72.15  & 0.37  & 40.00  & 5.85  & 1.18  & 0.43  \\
				& LP-BFGS+CE & 80.96 & \textbf{0.9} & 39.99 & 13.31 & 2.5   & 0.78  & \textbf{76.23} & \textbf{0.63} & 39.99  & 15.73  & 2.89  & 0.84  \\
				& LP-BFGS+CW LOG & \textbf{84.15} & 0.44  & 40    & 4.59  & 0.91  & 0.33  & 71.43  & 0.37  & 40.00  & 5.92  & 1.19  & 0.43  \\
				\multicolumn{1}{c}{k=40} & FGSM  & 35.92 & 0.69  & 39.94 & 20.45 & 3.55  & 0.87  & 40.22  & 0.48  & 39.93  & 20.83  & 3.61  & 0.88  \\
				\multicolumn{1}{c}{$\lambda=1.6$} & C\&W  & 82.9  & 0.46  & 40    & 2.86  & 0.54  & 0.18  & 75.15  & 0.37  & 40.00  & 4.83  & 0.89  & 0.26  \\
				& SparseFool & 49.17 & 0.64  & 2223.18 & 1117.98 & 22.54 & 0.8   & 74.79  & 0.54  & 1602.39  & 805.82  & 16.68  & 0.81  \\
				& JSMA  & 51.27 & 0.57  & 26.12 & 13.74 & 2.77  & 0.82  & 50.42  & 0.43  & 24.59  & 13.27  & 2.75  & 0.82  \\
				\midrule
				& LP-BFGS+CW & 93.02 & 0.44  & 60    & 6.7   & 1.1   & 0.34  & 85.23  & 0.36  & 60.00  & 7.02  & 1.17  & 0.37  \\
				& LP-BFGS+CE & 89.86 & 0.92  & 59.99 & 20.58 & 3.15  & 0.81  & 88.36  & \textbf{0.70} & 60.00  & 23.06  & 3.47  & 0.85  \\
				& LP-BFGS+CW LOG & \textbf{93.24} & 0.44  & 60    & 6.55  & 1.08  & 0.34  & 85.11  & 0.36  & 60.00  & 7.05  & 1.18  & 0.37  \\
				\multicolumn{1}{c}{k=60} & FGSM  & 43.13 & 0.69  & 59.9  & 30.63 & 4.34  & 0.89  & 55.07  & 0.48  & 59.87  & 30.97  & 4.37  & 0.89  \\
				\multicolumn{1}{c}{$\lambda=2.4$} & C\&W  & 90.43 & 0.47  & 60    & 3.68  & 0.57  & 0.15  & 88.74  & 0.36  & 60.00  & 5.60  & 0.84  & 0.20  \\
				& SparseFool & 82.91 & 0.66  & 2493.52 & 1254.76 & 25.47 & 0.87  & \textbf{91.24} & 0.54  & 1667.14  & 841.23  & 17.45  & 0.83  \\
				& JSMA  & 65.59 & 0.57  & 34.57 & 17.6  & 3.08  & 0.83  & 69.59  & 0.43  & 31.72  & 16.85  & 3.05  & 0.83  \\
				\midrule
				& LP-BFGS+CW & \textbf{96.96} & 0.44  & 80    & 7.27  & 1.05  & 0.3   & 90.87  & 0.37  & 80.00  & 8.32  & 1.23  & 0.36  \\
				& LP-BFGS+CE & 94.93 & \textbf{0.94} & 79.98 & 26.86 & 3.57  & 0.82  & 94.17  & \textbf{0.77} & 79.99  & 31.04  & 4.04  & 0.87  \\
				& LP-BFGS+CW LOG & 96.84 & 0.44  & 80    & 7.16  & 1.03  & 0.3   & 89.73  & 0.36  & 80.00  & 8.45  & 1.24  & 0.36  \\
				\multicolumn{1}{c}{k=80} & FGSM  & 46.9  & 0.67  & 79.72 & 40.97 & 5.02  & 0.89  & 56.27  & 0.50  & 79.78  & 41.36  & 5.05  & 0.90  \\
				\multicolumn{1}{c}{$\lambda=3.2$} & C\&W  & 96.05 & 0.48  & 80    & 4.08  & 0.54  & 0.12  & \textbf{94.42} & 0.37  & 80.00  & 6.82  & 0.89  & 0.19  \\
				& SparseFool & 91.67 & 0.68  & 2547.48 & 1283.55 & 26.06 & 0.89  & 93.28  & 0.55  & 1977.57  & 996.38  & 20.45  & 0.86  \\
				& JSMA  & 76.58 & 0.54  & 42.1  & 20.64 & 3.29  & 0.84  & 75.41  & 0.44  & 40.07  & 20.83  & 3.39  & 0.84  \\
				\midrule
				& LP-BFGS+CW & 97.98 & 0.44  & 100   & 8.66  & 1.13  & 0.31  & 90.05  & 0.36  & 100.00  & 8.88  & 1.17  & 0.32  \\
				& LP-BFGS+CE & 96.72 & \textbf{0.95} & 99.99 & 33.15 & 3.94  & 0.82  & \textbf{97.45} & \textbf{0.81} & 100.00  & 38.33  & 4.48  & 0.87  \\
				& LP-BFGS+CW LOG & 98.32 & 0.44  & 100   & 8.47  & 1.11  & 0.3   & 89.93  & 0.36  & 100.00  & 8.81  & 1.17  & 0.32  \\
				\multicolumn{1}{c}{k=100} & FGSM  & 48.88 & 0.66  & 99.84 & 50.95 & 5.58  & 0.9   & 61.77  & 0.50  & 99.76  & 51.19  & 5.60  & 0.91  \\
				\multicolumn{1}{c}{$\lambda=4.0$} & C\&W  & \textbf{98.54} & 0.49  & 100   & 4.8   & 0.58  & 0.12  & 96.97  & 0.37  & 100.00  & 7.36  & 0.85  & 0.16  \\
				& SparseFool & 95.99 & 0.68  & 2612.46 & 1314.72 & 26.75 & 0.9   & 96.60  & 0.56  & 2100.33  & 1057.29  & 21.64  & 0.88  \\
				& JSMA  & 85.84 & 0.55  & 47.39 & 23.16 & 3.47  & 0.84  & 84.34  & 0.44  & 41.95  & 21.57  & 3.41  & 0.85  \\
				\bottomrule
			\end{tabular}%
		}
		\label{tab:results-on-cifar10}%
	\end{table*}

	\section{Experiments}
	
	\begin{table*}[!t]
		\centering
		\caption{The performance of LP-BFGS (with 3 loss functions), FGSM \cite{FGSM}, C\&W \cite{C&W} on the ImageNet dataset. The target models are VGG-19 \cite{VGG} and Res-34 \cite{ResNet}.}
		\resizebox{\textwidth}{!}{
			\begin{tabular}{llcccccccccccc}
				\toprule
				&       & \multicolumn{6}{c}{VGG-19}                    & \multicolumn{6}{c}{Res-34} 
				\\
				\cmidrule(l){3-8} \cmidrule(l){9-14}  
				\multicolumn{1}{c}{Pixel} & Attack & ASR (\%) & Confidence & \multicolumn{1}{l}{$L_0$} & \multicolumn{1}{l}{$L_1$} & \multicolumn{1}{l}{$L_2$} & \multicolumn{1}{l}{$L_{\infty}$} & ASR (\%) & Confidence & \multicolumn{1}{l}{$L_0$} & \multicolumn{1}{l}{$L_1$} & \multicolumn{1}{l}{$L_2$} & \multicolumn{1}{l}{$L_{\infty}$} 
				\\
				\cmidrule{1-14}          & LP-BFGS+CW & 9.71  & 0.23  & 20    & 2.58  & 0.73  & 0.35  & 14.58 & 0.24  & 20    & 3.5   & 0.97  & 0.44 \\
				& LP-BFGS+CE & \textbf{14.57} & \textbf{0.27} & 19.99 & 6.82  & 1.83  & 0.78  & \textbf{15.06} & \textbf{0.32} & 20    & 7.63  & 1.99  & 0.78 \\
				\multicolumn{1}{c}{k=20} & LP-BFGS+CW LOG & 9.86  & 0.22  & 20    & 2.56  & 0.73  & 0.35  & 13.31 & 0.24  & 20    & 3.38  & 0.95  & 0.43 \\
				& FGSM  & 7.71  & 0.26  & 20    & 10.62 & 2.6   & 0.87  & 8.56  & 0.3   & 19.98 & 10.49 & 2.56  & 0.85 \\
				& C\&W  & 12.43 & 0.22  & 20    & 2.14  & 0.54  & 0.2   & 14.1  & 0.25  & 20    & 2.79  & 0.71  & 0.27 \\
				\midrule
				& LP-BFGS+CW & 21.38 & 0.21  & 40    & 5.82  & 1.16  & 0.41  & 22.63 & 0.24  & 40    & 6.75  & 1.34  & 0.46 \\
				& LP-BFGS+CE & \textbf{26.69} & \textbf{0.26} & 40    & 13.63 & 2.58  & 0.81  & \textbf{24.46} & \textbf{0.33} & 40    & 14.29 & 2.68  & 0.82 \\
				\multicolumn{1}{c}{k=40} & LP-BFGS+CW LOG & 22.81 & 0.21  & 40    & 5.88  & 1.17  & 0.42  & 22.17 & 0.24  & 40    & 6.8   & 1.34  & 0.46 \\
				& FGSM  & 12.2  & 0.25  & 40    & 21.32 & 3.68  & 0.9   & 11.62 & 0.31  & 39.95 & 21.07 & 3.66  & 0.9 \\
				& C\&W  & 23.96 & 0.21  & 40    & 4.55  & 0.84  & 0.25  & 22.32 & 0.24  & 40    & 5.09  & 0.93  & 0.27 \\
				\midrule
				& LP-BFGS+CW & 28.8  & 0.21  & 60    & 7.72  & 1.29  & 0.42  & 29.97 & 0.26  & 60    & 8.92  & 1.48  & 0.45 \\
				& LP-BFGS+CE & \textbf{33.04} & \textbf{0.28} & 60    & 19.27 & 3.02  & 0.83  & \textbf{35.02} & \textbf{0.37} & 60    & 21.75 & 3.36  & 0.86 \\
				\multicolumn{1}{c}{k=60} & LP-BFGS+CW LOG & 28.65 & 0.21  & 60    & 7.79  & 1.3   & 0.42  & 28.75 & 0.26  & 60    & 8.85  & 1.47  & 0.45 \\
				& FGSM  & 14.04 & 0.23  & 60    & 31.31 & 4.47  & 0.91  & 16.67 & 0.31  & 59.94 & 30.85 & 4.43  & 0.91 \\
				& C\&W  & 31.29 & 0.22  & 60    & 6.18  & 0.94  & 0.25  & 32.72 & 0.26  & 60    & 7.13  & 1.09  & 0.28 \\
				\midrule
				& LP-BFGS+CW & 32.79 & 0.22  & 80    & 9.34  & 1.36  & 0.39  & 36.03 & 0.24  & 80    & 11.39 & 1.64  & 0.45 \\
				& LP-BFGS+CE & \textbf{40.24} & \textbf{0.32} & 80    & 25.09 & 3.46  & 0.84  & \textbf{42.6} & \textbf{0.36} & 80    & 28.03 & 3.76  & 0.86 \\
				\multicolumn{1}{c}{k=80} & LP-BFGS+CW LOG & 32.19 & 0.22  & 80    & 9.18  & 1.34  & 0.39  & 34.5  & 0.24  & 80    & 11.14 & 1.6   & 0.44 \\
				& FGSM  & 16.39 & 0.25  & 79.99 & 41.41 & 5.16  & 0.91  & 20.76 & 0.28  & 79.85 & 40.85 & 5.04  & 0.91 \\
				& C\&W  & 36.96 & 0.23  & 80    & 7.16  & 0.94  & 0.21  & 37.25 & 0.25  & 80    & 8.17  & 1.08  & 0.24 \\
				\midrule
				& LP-BFGS+CW & 40.2  & 0.23  & 100   & 12.81 & 1.68  & 0.45  & 44.55 & 0.26  & 100   & 14.23 & 1.83  & 0.46 \\
				& LP-BFGS+CE & \textbf{47.98} & \textbf{0.33} & 100   & 30.94 & 3.81  & 0.85  & \textbf{52.23} & \textbf{0.44} & 100   & 35.15 & 4.21  & 0.87 \\
				\multicolumn{1}{c}{k=100} & LP-BFGS+CW LOG & 40.35 & 0.24  & 100   & 13    & 1.69  & 0.44  & 44.39 & 0.26  & 100   & 13.6  & 1.76  & 0.45 \\
				& FGSM  & 18.44 & 0.27  & 99.95 & 51.75 & 5.68  & 0.9   & 23.5  & 0.33  & 99.95 & 50.99 & 5.68  & 0.92 \\
				& C\&W  & 42.22 & 0.23  & 100   & 9.67  & 1.16  & 0.25  & 47.93 & 0.27  & 100   & 10.56 & 1.25  & 0.26 \\
				\bottomrule
			\end{tabular}%
		}
		\label{tab:results-on-imagenet}
	\end{table*}
	
	\begin{figure*}
		\begin{minipage}[t]{1.0\linewidth}
			\centering
			\captionof{table}{The time cost of SparseFool \cite{SparseFool} on the ImageNet dataset. The target model is Res-34 \cite{ResNet}. Note that due to its high complexity, SparseFool is evaluated on only 100 examples.}
			\label{tab:time-compare}
			\resizebox{0.8 \linewidth}{!}{
				\begin{tabular}{cccccccc}
					\toprule
					\multicolumn{1}{l}{$\lambda$} & ASR (\%) & Confidence & \multicolumn{1}{l}{$L_0$} & \multicolumn{1}{l}{$L_1$} & \multicolumn{1}{l}{$L_2$} & \multicolumn{1}{l}{$L_{\infty}$} & \multicolumn{1}{c}{Time(ms)} \\
					\midrule
					8.0     & 97.18 & 0.32  & 128150.2 & 65453.9 & 196.53 & 0.97  & 385739.9 \\
					16.0    & 100   & 0.31  & 137884.7 & 69553.41 & 206.5 & 0.98  & 3242.57 \\
					\bottomrule
			\end{tabular}}
		\end{minipage}
		
		\begin{minipage}[t]{\textwidth}
			\centering
			\includegraphics[width=0.6 \linewidth]{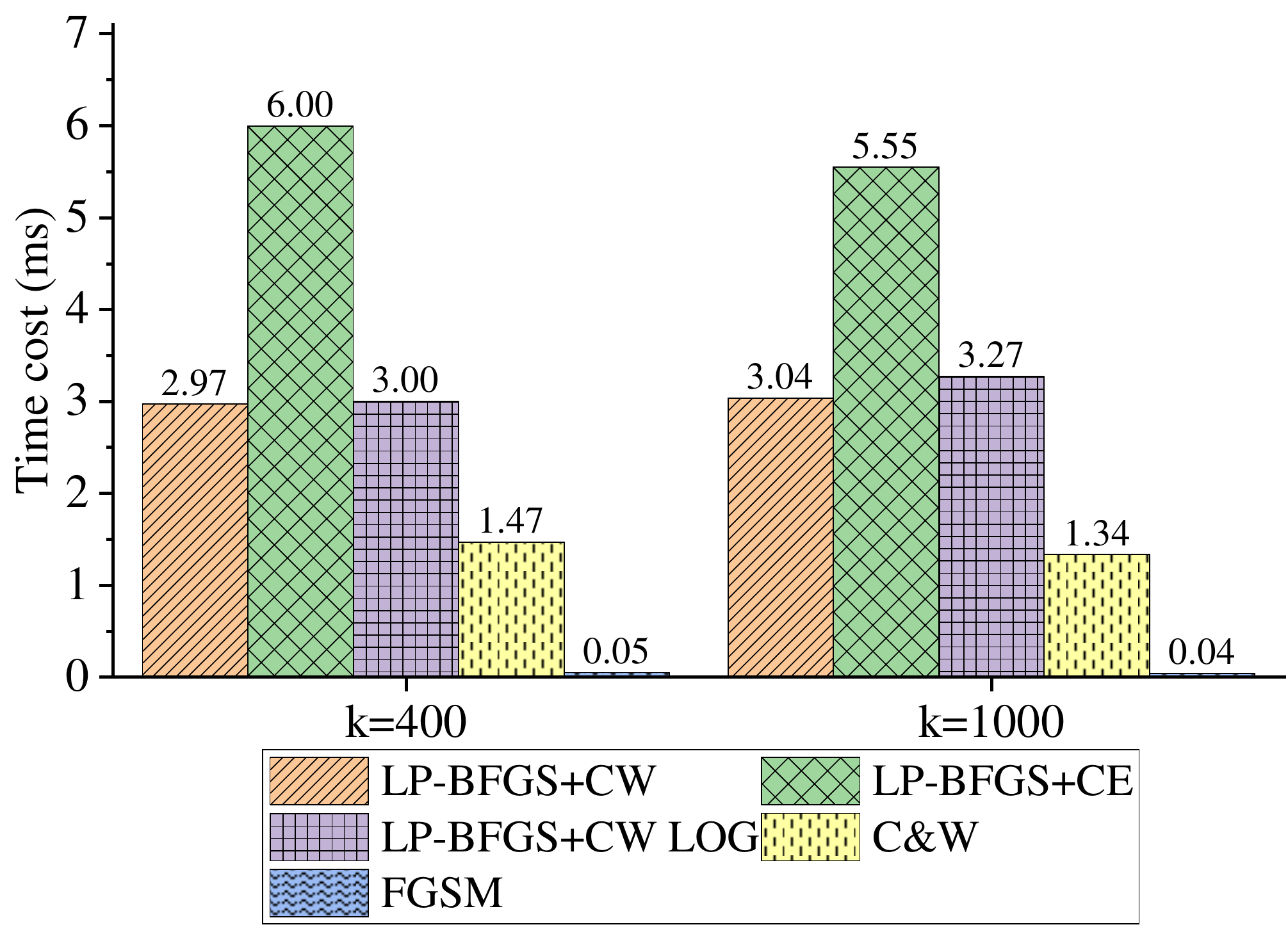}
			\caption{The time cost comparison with the perturbed pixel number $k=400$ and $k=1000$ on the ImageNet dataset. The target model is Res-34 \cite{ResNet}.}
			\label{fig:time-compare}
		\end{minipage}
	\end{figure*}

	\subsection{Setting}
	\subsubsection{Datasets and models}
	We conduct experiments based on CIFAR-10 \cite{CIFAR-10} and ImageNet \cite{ImageNet} datasets, utilizing pre-trained models as target models. The pre-training models on the CIFAR-10 dataset are Res-20 \cite{ResNet} and NIN \cite{NIN}, while on the ImageNet dataset, they are VGG-19 \cite{VGG} and Res-34 \cite{ResNet}. 
	\subsubsection{Attacks and metrics}
	We consider FGSM \cite{FGSM}, C\&W \cite{C&W}, JSMA \cite{JSMA}, and SparseFool \cite{SparseFool} attack as baselines in this work.
	It is worth noting that JSMA is not evaluated on ImageNet dataset, due to its huge computational cost for searching over all pairs of candidates \cite{C&W}.
	The LP-BFGS with three loss functions (expressed in \eqref{eq:CE_loss}, \eqref{eq:CW_loss}, and \eqref{eq:CW_log_loss}) are denoted by LP-BFGS+CE, LP-BFGS+CW, LP-BFGS+CW LOG, respectively. Since the perturbed pixels have a significant impact on the attack effectiveness, the optimization space of FGSM, C\&W, and LP-BFGS is limited to the pixels selected by the IG. For JSMA and SparseFool, the optimization space is not restricted. The iterations dominate the number of perturbed pixels in JSMA attack and SparseFool uses the hyperparameter $\lambda$ to govern the number of perturbed pixels. 
	For each evaluation, 1000 images are randomly selected from the test as the original examples. In addition to the attack success rate (ASR) and confidence, $L_1$, $L_2$, and $L_{\infty}$ norms of the perturbation vector are evaluated to measure the perturbation magnitude.
	\par
	The perturbation generated by FGSM is only constrained by the $L_{\infty}$ norm. To achieve a strong attack, the $L_{\infty}$ norm of perturbation is empirically set to 1 in FGSM.
	For the C\&W attack, the $L_2$ attack based on the pixel selector is conducted, $c$ regulating the perturbations size and attack ability is set to $10^3$, $\kappa$ is set to $0$, and the iterations is 200. The Adam optimizer's step size is set to 0.1.
	For the LP-BFGS attack, $c$, $\kappa$, and iterations are consistent with the C\&W attack, and three loss functions designed in Section \ref{sect:lp-bfgs} are adopted.
	For SparseFool, the hyperparameter $\lambda$ controls perturbed pixel number and attack effectiveness. Generally, SparseFool has a stronger attack performance and more perturbed pixels with a growing $\lambda$. The iteration of SparseFool is consistent with the C\&W attack. The $L_0$ distance of perturbations reflects the average number of perturbed pixels.
	JSMA attack is a targeted attack method that modifies two pixels per iteration. The iteration determines the maximal number of perturbed pixels. Therefore, the iteration of the JSMA attack is set to half of the perturbation budget. The perturbation size of JSMA is also set to 1. We randomly select a label (different from the true label) as the target one for executing the JSMA attack, since we mainly explore the performance of untargeted attacks.
	All experiments are conducted on a machine with an Intel(R) Core(TM) i7-9700 CPU, 32 GB RAM, and a single NVIDIA GeForce RTX 2080 Ti GPU.
	
	\begin{table*}[!t]
		\centering
		\caption{The performance of LP-BFGS (with 3 loss functions), FGSM \cite{FGSM}, C\&W \cite{C&W} on the ImageNet dataset with different pixel selectors. The target model is Res-34  \cite{ResNet}.}
		\label{tab:selector-compare}%
		\resizebox{0.8 \textwidth}{!}{
			\begin{tabular}{c|c|lcccccc}
				\toprule
				\multicolumn{1}{c}{Strategy} & \multicolumn{1}{c}{Pixel} & Attack & ASR(\%) & Confidence & \multicolumn{1}{c}{$L_0$} & \multicolumn{1}{c}{$L_1$} & \multicolumn{1}{c}{$L_2$} & \multicolumn{1}{c}{$L_{\infty}$} \\
				\midrule
				& \multicolumn{1}{c|}{40} & LP-BFGS+CW & 22.63  & 0.24  & 40.00  & 6.75  & 1.34  & 0.46  \\
				& \multicolumn{1}{c|}{40} & LP-BFGS+CE & \textbf{24.46}  & 0.33  & 40.00  & 14.29  & 2.68  & 0.82  \\
				& \multicolumn{1}{c|}{40} & LP-BFGS+CW LOG & 22.17  & 0.24  & 40.00  & 6.80  & 1.34  & 0.46  \\
				& \multicolumn{1}{c|}{40} & FGSM  & 11.62  & 0.31  & 39.95  & 21.07  & 3.66  & 0.90  \\
				& \multicolumn{1}{c|}{40} & C\&W  & 22.32  & 0.24  & 40.00  & 5.09  & 0.93  & 0.27  \\
				\cmidrule{2-9}      & \multicolumn{1}{c|}{400} & LP-BFGS+CW & 76.26  & 0.26  & 400.00  & 38.58  & 2.59  & 0.41  \\
				& \multicolumn{1}{c|}{400} & LP-BFGS+CE & \textbf{89.31}  & 0.70  & 400.00  & 139.07  & 8.38  & 0.92  \\
				IG+top-k & \multicolumn{1}{c|}{400} & LP-BFGS+CW LOG & 77.04  & 0.26  & 399.99  & 39.61  & 2.66  & 0.42  \\
				& \multicolumn{1}{c|}{400} & FGSM  & 44.18  & 0.30  & 399.27  & 203.84  & 11.34  & 0.95  \\
				& \multicolumn{1}{c|}{400} & C\&W  & 85.06  & 0.27  & 400.00  & 29.53  & 1.77  & 0.20  \\
				\cmidrule{2-9}      & \multicolumn{1}{c|}{1000} & LP-BFGS+CW & 83.87  & 0.27  & 999.98  & 60.90  & 2.69  & 0.33  \\
				& \multicolumn{1}{c|}{1000} & LP-BFGS+CE & \textbf{97.60}  & 0.93  & 1000.00  & 331.87  & 12.75  & 0.93  \\
				& \multicolumn{1}{c|}{1000} & LP-BFGS+CW LOG & 87.22  & 0.27  & 999.97  & 60.89  & 2.69  & 0.33  \\
				& \multicolumn{1}{c|}{1000} & FGSM  & 58.95  & 0.32  & 997.07  & 507.74  & 17.83  & 0.96  \\
				& \multicolumn{1}{c|}{1000} & C\&W  & 96.17  & 0.29  & 1000.00  & 48.88  & 1.86  & 0.13  \\
				\midrule
				& \multicolumn{1}{c|}{40} & LP-BFGS+CW & 8.32  & 0.27  & 40.00  & 6.98  & 1.59  & 0.65  \\
				& \multicolumn{1}{c|}{40} & LP-BFGS+CE & \textbf{9.09}  & 0.32  & 40.00  & 12.35  & 2.56  & 0.88  \\
				& \multicolumn{1}{c|}{40} & LP-BFGS+CW LOG & 8.32  & 0.25  & 40.00  & 6.85  & 1.53  & 0.62  \\
				& \multicolumn{1}{c|}{40} & FGSM  & 5.24  & 0.32  & 39.26  & 20.21  & 3.74  & 0.96  \\
				& \multicolumn{1}{c|}{40} & C\&W  & 7.24  & 0.25  & 40.00  & 4.61  & 0.98  & 0.36  \\
				\cmidrule{2-9}      & \multicolumn{1}{c|}{400} & LP-BFGS+CW & 55.56  & 0.28  & 399.99  & 53.74  & 3.94  & 0.68  \\
				& \multicolumn{1}{c|}{400} & LP-BFGS+CE & \textbf{63.81}  & 0.53  & 399.90  & 119.91  & 7.85  & 0.95  \\
				Random & \multicolumn{1}{c|}{400} & LP-BFGS+CW LOG & 54.65  & 0.28  & 399.99  & 54.61  & 4.00  & 0.68  \\
				& \multicolumn{1}{c|}{400} & FGSM  & 24.47  & 0.31  & 395.02  & 199.81  & 11.46  & 0.98  \\
				& \multicolumn{1}{c|}{400} & C\&W  & 60.66  & 0.29  & 400.00  & 42.96  & 2.81  & 0.36  \\
				\cmidrule{2-9}      & \multicolumn{1}{c|}{1000} & LP-BFGS+CW & 76.40  & 0.30  & 999.98  & 95.90  & 4.63  & 0.62  \\
				& \multicolumn{1}{c|}{1000} & LP-BFGS+CE & \textbf{94.70}  & 0.81  & 999.93  & 302.42  & 12.38  & 0.96  \\
				& \multicolumn{1}{c|}{1000} & LP-BFGS+CW LOG & 75.49  & 0.30  & 999.99  & 96.10  & 4.64  & 0.63  \\
				& \multicolumn{1}{c|}{1000} & FGSM  & 39.33  & 0.34  & 983.38  & 500.57  & 18.11  & 0.98  \\
				& \multicolumn{1}{c|}{1000} & C\&W  & 93.19  & 0.31  & 1000.00  & 90.05  & 3.64  & 0.29  \\
				\midrule
				& 40    & LP-BFGS+CW & 4.48  & 0.26  & 40.00  & 7.51  & 1.77  & 0.74  \\
				& 40    & LP-BFGS+CE & \textbf{5.40}  & 0.30  & 39.71  & 11.78  & 2.57  & 0.88  \\
				& 40    & LP-BFGS+CW LOG & 5.25  & 0.26  & 40.00  & 6.69  & 1.60  & 0.71  \\
				& 40    & FGSM  & 3.09  & 0.32  & 32.80  & 20.26  & 4.03  & 0.96  \\
				& 40    & C\&W  & 4.17  & 0.25  & 40.00  & 4.59  & 1.06  & 0.42  \\
				\cmidrule{2-9}      & 400   & LP-BFGS+CW & 36.06  & 0.28  & 399.98  & 54.53  & 4.36  & 0.76  \\
				& 400   & LP-BFGS+CE & \textbf{39.37}  & 0.45  & 399.38  & 104.15  & 7.48  & 0.96  \\
				IG+bottom-k & 400   & LP-BFGS+CW LOG & 35.91  & 0.28  & 400.00  & 56.13  & 4.48  & 0.78  \\
				& 400   & FGSM  & 13.39  & 0.29  & 366.93  & 202.09  & 12.34  & 0.98  \\
				& 400   & C\&W  & 37.80  & 0.28  & 400.00  & 36.77  & 2.67  & 0.42  \\
				\cmidrule{2-9}      & 1000  & LP-BFGS+CW & 55.52  & 0.31  & 1000.00  & 102.04  & 5.44  & 0.74  \\
				& 1000  & LP-BFGS+CE & \textbf{69.12}  & 0.68  & 999.91  & 252.40  & 11.46  & 0.97  \\
				& 1000  & LP-BFGS+CW LOG & 56.32  & 0.31  & 1000.00  & 103.02  & 5.46  & 0.74  \\
				& 1000  & FGSM  & 21.92  & 0.36  & 920.36  & 502.96  & 19.54  & 0.99  \\
				& 1000  & C\&W  & 68.32  & 0.32  & 1000.00  & 79.68  & 3.61  & 0.37  \\
				\bottomrule
			\end{tabular}
		}
	\end{table*}
	
	\begin{figure*}[!t]
		\centering
		\centerline{\includegraphics[width=1 \textwidth]{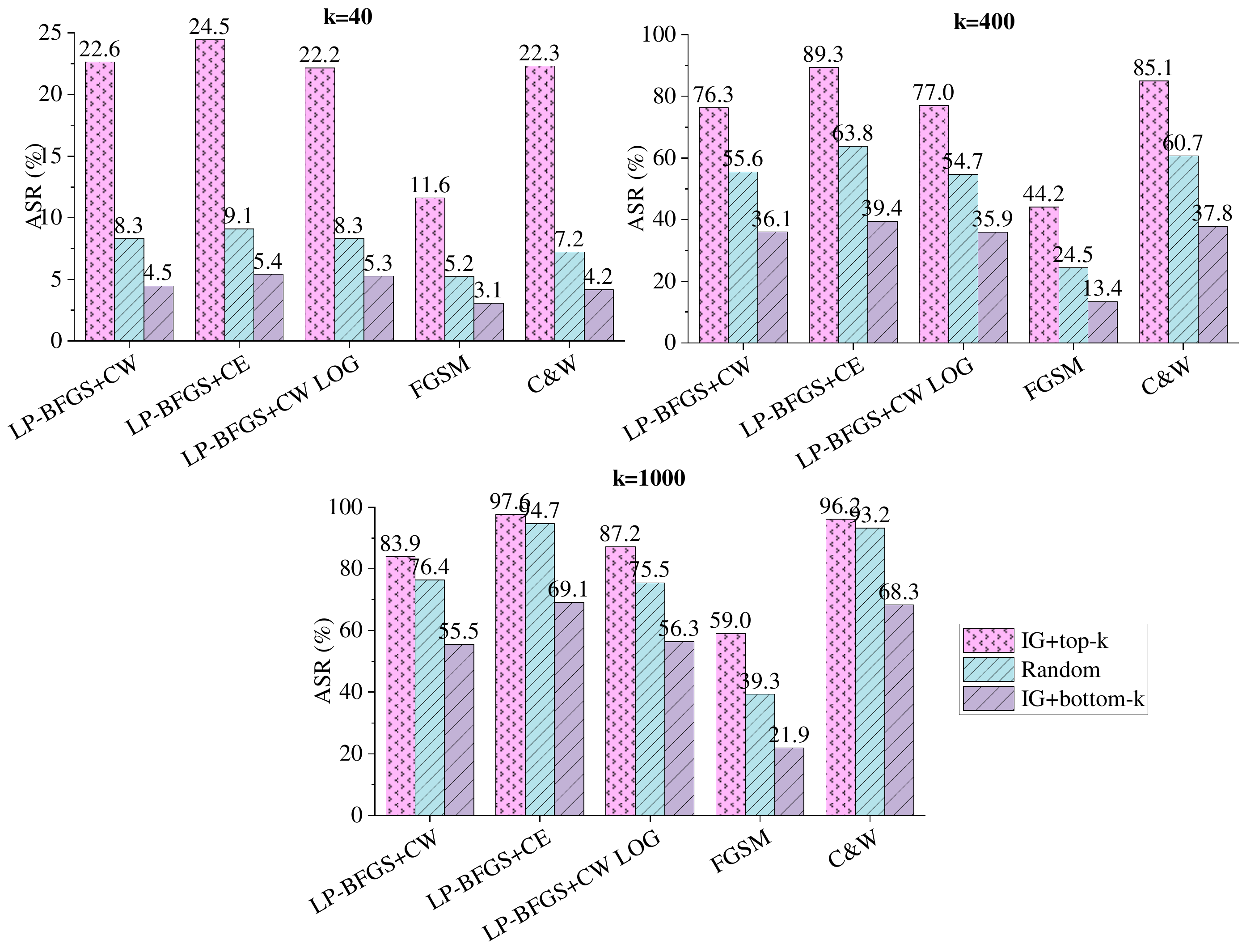}}
		\caption{The ASR comparison with the number of perturbed pixel $k$=40, 400, and 1000 on the ImageNet dataset. The target model is Res-34 \cite{ResNet}.}
		\label{fig:stragety-compare}
	\end{figure*}
	\subsection{Results with different numbers of perturbed pixels}
	We compare the attack performance of LP-BFGS, FGSM, C\&W, JSMA, and SpareFool with the different numbers of perturbed pixels, and the detailed evaluation results are shown in Table~\ref{tab:results-on-cifar10}, Table~\ref{tab:results-on-imagenet}, and Table~\ref{tab:selector-compare}. Fig.~\ref{fig:adv-images} shows some adversarial images generated by the attack methods mentioned above. Since SpareFool achieves a compromise between perturbed pixels and the attack effect by $\lambda$, the values of $\lambda$ are shown in Table~\ref{tab:results-on-cifar10}.
	The results from the tables show that the ASR of all attack methods overall increases as the number of perturbed pixels increases, but compared to FGSM, JSMA, SpareFool, and C\&W, LP-BFGS generally has a higher value.
	On the CIFAR-10 dataset, LP-BFGS slightly outperforms C\&W in terms of ASR in most cases. On the ImageNet dataset, LP-BFGS has a prominent advantage.
	\par
	The attack performance of FGSM on the CIFAR-10 and ImageNet dataset ranks at the bottom, while this gap is highlighted in complicated classification tasks, \textit{e.g.}, when attacking the model VGG-19 trained on the ImageNet dataset with the number of perturbed pixels k= 100, FGSM only obtains 18.44\% ASR, which lags behind the 47.98\% of LP-BFGS+CE.
	\par
	
	SparseFool shows excellent attack performance on the CIFAR-10 dataset with a sufficient perturbation budget, but there is still a gap with the LP-BFGS attack. Furthermore, SparseFool could implicitly regulate the perturbation budget but needs more perturbation budget than that of LP-BFGS when it achieves a comparable attack performance.
	when $\lambda$ is 1.0 and 4.0, the average number of perturbed pixels of SparseFool is 459 and 2612, respectively, while LP-BFGS only perturbs 100 pixels.
	
	\par
	JSMA also exhibits excellent attack performance, but it can be seen that JSMA does not fully utilize the perturbation budget. The average $L_0$ norm of perturbations is lower than the preset value, which may be the reason for the weak performance of JSMA compared to LP-BFGS, C\&W, and SparseFool.

	\subsection{Results with different loss functions}
	In this section, we explore the effectiveness of the three loss functions (namely CE, CW, and CW LOG) described in Section~\ref{sect:lp-bfgs}. As can be seen from Table~\ref{tab:results-on-cifar10}, CW LOG generally outperforms the remaining two loss functions on the CIFAR-10 dataset. CE has the weakest attack capability among the three loss functions. However, the situation goes into reverse on the ImageNet dataset. CE is more helpful to improve the attack performance compared with the other two ones. The reason behind this pronounced contrast may be that the CE can amplify the difference between the true label and other labels, which is more conducive to the optimization process. In addition, the confidence of adversarial examples generated by CE has the highest value on all datasets and models. 
	CW and CW LOG are very similar with respect to the attack performance and perturbation magnitude.
	Compared with CE, they are inclined to generate stealthy perturbations.
	
	\subsection{Results with different pixel selection strategies}
	To investigate the effect of the pixel selection strategy incorporated by LP-BFGS, besides pixels with top-k attribution scores as perturbation pixels, we design the other two selection strategies, \textit{i.e.}, randomly selected pixels as perturbation pixels and pixels with bottom-k attribution scores as perturbation pixels. These strategies are denoted as IG+top-k, Random, and IG+bottom-k, respectively. Table~\ref{tab:selector-compare} reports the evaluation results of LP-BFGS based on different selection strategies. Fig.~\ref{fig:stragety-compare} visualizes the ASR derived by different strategies.
	It can be seen that IG + top-k overall has a obvious improvement on the attack performance of all attacks in contrast with the others.
	Moreover, IG+top-k has a significant advantage compared with Random and IG+bottom-k when we have an inadequate perturbation budget. But the gap is narrowing as the number of perturbed pixels increases. When the number of perturbed pixels reaches 1000, the ASR of LP-BFGS+CE is 97.6\%, 94.7\%, and 69.12\% for the three strategies, respectively. This shows that for an image of size $3\times256\times256$, the perturbed pixels that account for only 0.51\% of the total number of pixels can cause a strong attack against the target model. Not only adding a small perturbation to the input can significantly degrade the performance of the model, but changing only a small fraction of the input can mislead the prediction of the model.

	\subsection{Results of the time consumption}
	Table~\ref{tab:time-compare} and Fig.~\ref{fig:time-compare} show the average execution time of LP-BFGS, C\&W, SparseFool, and FGSM for the Res-34 model trained on ImageNet. When $\lambda$ = 4.0, the ASR reaches 97.18\%, the average number of perturbed pixels is 128150, which accounts for 65.18\% of total pixels, and the average execution time is 385739.9 ms. Table~\ref{tab:selector-compare} shows that the ASRs of LP-BFGS+CE and C\&W are 97.60\% and 96.17\%, with 1000 perturbed pixels, respectively. The average attack time is 5.55 ms and 1.34 ms, respectively.
	These results show that SparseFool not only has a larger perturbation magnitude but also has a higher time consumption while it achieves a comparable attack. When there is a large size image, the search space is vast and SparseFool needs to spend more time to find sparse perturbations in the LinearSolver. From Table~\ref{tab:selector-compare} and Fig.~\ref{fig:time-compare}, it can be seen that although LP-BFGS achieves an advantage in the complicated classification tasks, the time cost is higher than that of the C\&W attack.
	
	\section{Conclusion}
	In this work, we find the sparsity requirement on the perturbation does reduce the size of the Hessian matrix and propose a $L_0$-norm based white box attack method, LP-BFGS. 
	We explore the performance of the proposed method from various aspects.
	Experimental results demonstrate our approach's outstanding attack ability in the limited perturbation pixel scenario compared with existing attacks. 
	But there is room for improvement in terms of the perceptibility of perturbations \cite{IS_3,IS_9}, and the time cost. Although the perturbation generated by LP-BFGS has a stronger attack ability, it has large $L_1$, $L_2$, and $L_{\infty}$ norms. 
	Furthermore, the selection of pixels, the design of loss functions \cite{EAD-attack}, and the usage of the optimization method are all worth exploring.

	
	

	
	\newpage
	\bibliographystyle{IEEEbib}
	\bibliography{lp-bfgs-refs}
	
\end{document}